\documentclass[aps,pre,showpacs,floats,twocolumn,floats,superscriptaddress,floatfix,groupedaddress]{revtex4-1}
\usepackage{bm}
\usepackage{url}
\usepackage{times}
\usepackage{verbatim}
\usepackage{theorem}
\usepackage{makeidx}
\usepackage{amsmath}
\usepackage{siunitx}
\usepackage{xy}
\usepackage{amssymb}
\usepackage{graphicx}
\usepackage{color}
\usepackage{epsfig}
\usepackage{subfigure}  
\usepackage[breaklinks]{hyperref}

\begin{document}
\title{Heavy inertial particles in rotating turbulence : distribution of particles in flow and evolution of Lagrangian trajectories}
\author{Priyanka Maity}
\email{priyanka.maity@tu-ilmenau.de}
\affiliation{Institute of Thermodynamics and Fluid Mechanics, Technische Universit\"at Ilmenau, Postfach 100565, D-98684 Ilmenau, Germany} 
\begin{abstract}

We revisited the problem of  heavy particles suspended in homogeneous box turbulence flow subjected to rotation along the vertical axis, which introduces anisotropy along the vertical and horizontal planes. We investigate the effect of the emergent structures due to rotation, on the spatial distribution and temporal statistics of the particles. The spatial distributions were studied using the joint probability distribution function (JPDFs) of the two invariants, $Q$ and $R$, of the velocity gradient tensor. At high rotation rates, the JPDFs of Lagrangian $Q-R$ plots show remarkable deviations from the well known \textit{teardrop} shape. The cumulative probability distribution functions (CDFs) for times during which a particle remains in vortical or straining regions, show exponentially decaying tails except for the deviations at the highest rotation rate. The average residence times of the particles in vortical and straining regions are also affected considerably due to the addition of rotation. In addition, we compute the temporal velocity autocorrelation and connect it to the Lagrangian anisotropy in presence of rotation. The spatial and temporal statistics of the particles are determined by a complex competition between the rotation rate and the heaviness of the particles.

\end{abstract}

\maketitle
\section{Introduction}
Turbulent flows in a rotating frame of reference~\cite{Greenspan_1968, Zhou_1995, cambon_1997, davidson_2013} are ubiquitous in nature and are observed in 
geophysical~\cite{Aurnou_2015}, astrophysical~\cite{Barnes_2001,James_2008,Reun_2017} and industrial systems~\cite{Dumitrescu_2004}. Translation to a rotating frame of reference  gives rise to two pseudoforces in the system: the Coriolis and the centrifugal forces. The pseudoforces do not 
pump any additional energy into the turbulent systems, nevertheless they do make the turbulent flow anisotropic in nature. The anisotropy is generated due to an 
accumulation of energy in modes perpendicular to the plane of rotation~\cite{Smith_1999, Muller_2007, Minnini_2009, manohar_2019}, which eventually leads to 
two dimensionalization of the flow by generating columnar vortices parallel to the axis of rotation~\cite{Proudman_1916, Taylor_1917, Greenspan_1968, davidson_2013, 
Luca_2016}, in the rapid rotation limit. A constant angular rotation $ \Omega$ also introduces a characteristic timescale $\tau_{\Omega} \sim 1/\Omega$ and consequently, 
an important characteristic wave number known as the Zeman wave number $k_{\Omega}$ (wave number at which the inertial turnover time and $\tau_{\Omega}$ becomes comparable ) 
in the system. In the rapid rotation limit, a steeper scaling of energy spectra $E(k) \propto k^{-2}$ is observed for $k < k_{\Omega}$; whereas the typical scaling of 
$E(k) \propto k^{-5/3}$ is preserved for $k \geq k_{\Omega}$. An anisotropic enhancement of cyclonic vortex stretching~\cite{hopfinger_1982, Morize_2005, Gallet_2014, 
bartello_1994, sreenivasan_2008}, emergent inverse cascade of energy in 3D turbulence~~\cite{Bartello_1995,metais_1996, Yarom_2013} and suppression of intermittency in 
turbulence~\cite{Luca_2016,rathor_2019} are some unique attributes of rotating turbulence. {{Due to the profound effects of background rotation, which provide an idealistic environment to introduce isotropy without disturbing the total energy of the system, rotating turbulence have been extensively 
investigated in the Eulerian framework.}} However, the description of rotating turbulence from a Lagrangian point of view is recent and in early phases of 
development~\cite{Castello_2011, Castello_PRL_2011, Castello_PRE_2011, Luca_2016, Maity_2019, Maity_2020}. 

Lagrangian description of isotropic and homogeneous turbulence has been the topic of active research only for the past few decades
~\cite{Elghobashi_1994, Balkovsky_2001, mordant_2001, mordant_2004, Bec_2007_PRL, bec_2010, perlekar_2011, Bhatnagar_2016, bhatnagar_2018, Cencini_2017, Jucha_2014, Xu_2014, picardo_2020}, 
due to its relevance to the problem of particles in turbulence~\cite{maxey_riley_1983, maxey_1987, Toschi_2009, Bec_2014, saw_2014}. Heavy and lighter particles react anisotropically to the flow structures. Vortical structures act as repellers, ejecting out heavier particles due to the centrifugal force, leading to the phenomenon of \textit{ preferential concentration}~\cite{eaton_1994,Bec_2005,Bec_2007_pre,Bec_2007_PRL}. In the dissipative range, clustering of heavy particles is attributed to convergence of particles trajectories towards a dynamically evolving attractor with fractal dimension~\cite{Bec_2005, Bec_2007_pre, Bec_2007_physicaD, Bec_2007_PRL}. Consequently, the temporal distribution 
of particles in vortical and straining regions also shows asymmetry. Perlekar \textit{et al.}~\cite{perlekar_2011} showed that, in two-dimensional turbulence, the 
tails of cumulative distribution functions (CDFs) of persistence times of particles vary as power law and exponential function in case of vortical and straining 
region, respectively. However, in the case of 3D turbulence, CDFs of persistence time of both vortical and straining regions show an exponential decay~\cite{Bhatnagar_2016}. Another important perspective about the emergent structures, in presence of rotation, can be obtained by studying the velocity autocorrelation function in the 
Lagrangian framework. Previous studies~\cite{mordant_2001,mordant_2004,Castello_PRE_2011,yeung_pope_1989, Marko_Virant_1997} have shown that the velocity autocorrelations decay exponentially as a function of time t as ${ \exp} (- t/\tau_0$), evolving with a characteristic time comparable to the energy injection scale.

In this article, we investigate how the anisotropy emerging in rotating turbulence affects the spatial distribution and characteristic time scales associated with the 
Lagrangian particles. The paper is organized in the subsequent sections. Section II describes the Eulerian and Lagrangian equations describing the ambient fluid and 
suspended heavy particles, respectively. The physical significance of the parameters used in the simulations are also described in this section. Section III describes the results obtained by analyzing the spatial and the temporal statistics of the particles. The spatial statistics of the particles are probed by computing the Eulerian and Lagrangian invariants of the velocity-gradient tensor. We also calculated the cumulative distribution functions of the persistence times of particles in vortical or straining regions, to investigate the modifications in temporal statistics due to the presence of rotation.  {{The  Lagrangian velocity autocorrelation functions, computed distinctively for velocity components parallel and perpendicular to the axis of rotation, shows the effect of anisotropy on the massive particles. Section IV concludes with the summary of our investigations and enlists future prospects in the same direction.}}

\section{Governing equations and Simulations }

\subsection{Eulerian fluid fields}
The momentum equation for an anisotropic turbulent flow, is represented by the three-dimensional incompressible Navier-Stokes equation in a rotating frame of reference, for a fluid with constant density and velocity field ${\bf u}$ with kinematic viscosity $\nu$ and can be written as:

\begin{eqnarray}
\frac{\partial {\bf u}}{\partial t}+({\bf u}\cdot\nabla){\bf u} + 2 ({ \rm \mathbf {\Omega}} \times {\bf u}) &=& -\nabla P' +\nu \nabla^2{\bf u}+ {\bf f}\label{ns}\\
\nabla\cdot {\bf u} &=& 0\label{cont}
\end{eqnarray}
The anisotropy in the system emerges as a consequence of a solid body rotation, rotating with a constant angular velocity ${\rm \mathbf {\Omega}} \equiv (0,0,\rm \Omega)$, 
along the z-axis. The Navier-Stokes Eqs.~\ref{ns}-\ref{cont} in the rotating frame of reference were solved using an in-house code based on standard pseudospectral~\cite{canuto_book} 
method, in a triply periodic box of side $L = 2 \pi$ with a second order Adam-Bashforth scheme for time marching. Transitioning to rotating frame of reference, gives rise to pseudoforces viz., the Coriolis and the centrifugal force. The Coriolis force appears explicitly in the equation. However, the centrifugal contribution $({\rm \mathbf \Omega} \times {\rm \mathbf \Omega} \times \rm {\mathbf r})$, $\rm {\mathbf r}$ being the position vector of the fluid parcel from the center of the simulation box, is absorbed in the natural pressure term $P_0$, modifying it to $P' =  P_0 - \frac{1}{2} \vert {\rm {\mathbf \Omega}} \times {\bf r} \vert ^2$. This allows us to assume the periodic boundary condition, with the pressure getting modified as a result of the centrifugal force. A crucial step of turbulent simulations is to resolve the smallest length and time scales, viz., the Kolmogorov length $\eta$ and time scale $\tau_{\eta}$ respectively. For resolution of spatial scales, we used a grid resolution of 512 coallocation points in all the three directions which lead to $k_{max}\eta \approx 2.56$. The smallest Kolmogorov time  scale $\tau_{\eta}$ was resolved using a time stepping of $\delta_t = 4 \times 10^{-4}$, which ensured $\tau_{\eta}/\delta_t = 86$. To maintain a statistically steady state and counterbalance the loss due to viscous dissipation (with a mean dissipation rate of $\epsilon$),{ {we use a constant energy-injection external forcing ${\bf f}$, applied to the wavenumbers $k_f \leq 3$ (similar to the one used in reference~\cite{Sahoo_2011}).}} 

\begin{figure}
\begin{center}
\includegraphics[width=0.45\textwidth]{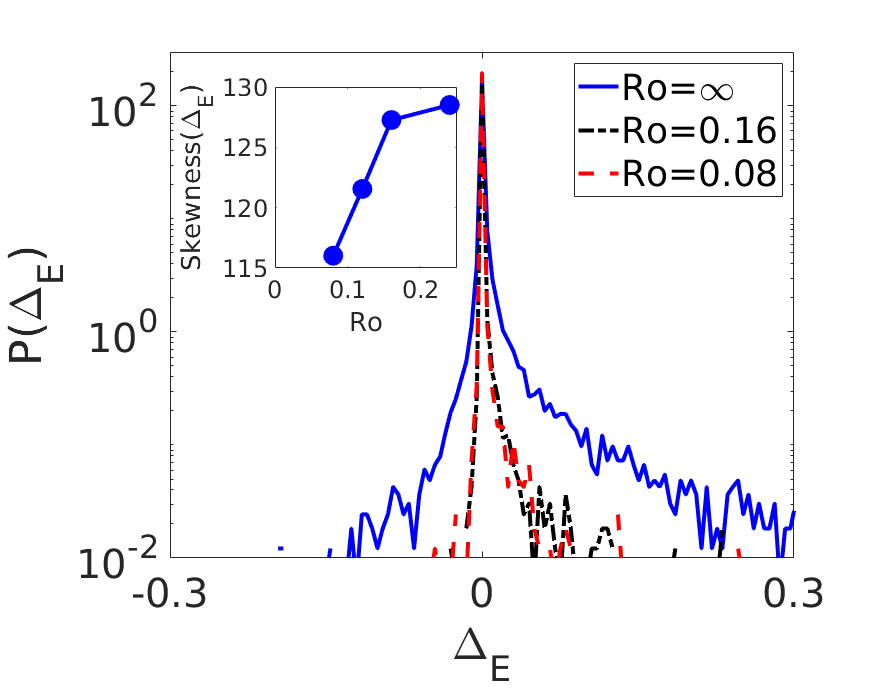}
\caption{Probability distribution functions (PDFs) of the discriminant of characteristic equation for the Eulerian velocity-gradient tensor $\Delta_E$ for three different strengths of rotation $Ro = \infty$, $0.16$, and $0.08$ (as denoted by the legends). { {The discriminant $\Delta_E$  has been nondimensionalized by the quantity $\chi^6$, where $\chi = u_{\eta}/\eta$}}. The inset shows skewness of the PDFs of $\Delta_E$ as a function of the Rossby number $\rm Ro$.}\label{Dpdf}
\end{center}
\end{figure}

The dynamics of the system are mainly governed by two dimensionless parameters : the Reynolds and the Rossby numbers. The dimensionless Reynolds number $Re ={{ u_{rms}} L}/{\nu}$ gives a ratio of inertial to viscous forces in the system, with $  u_{rms}$ denoting the root-mean-square velocity of the fluid field. In our simulations we use the Taylor-Reynolds number 
$Re_{\rm \lambda}$, the Reynolds number corresponding to the Taylor microscale $\lambda$, as a measure of turbulence. The Taylor-Reynolds number is defined as:

\begin{eqnarray}
Re_{\rm \lambda} &=& \frac{u_{\rm rms} \lambda}{\nu}\\
\lambda &=& \sqrt{\frac{15 \nu}{\epsilon}} u_{\rm rms}
\label{Re}
\end{eqnarray}

\begin{figure*}
\begin{center}
\includegraphics[width=0.9\textwidth]{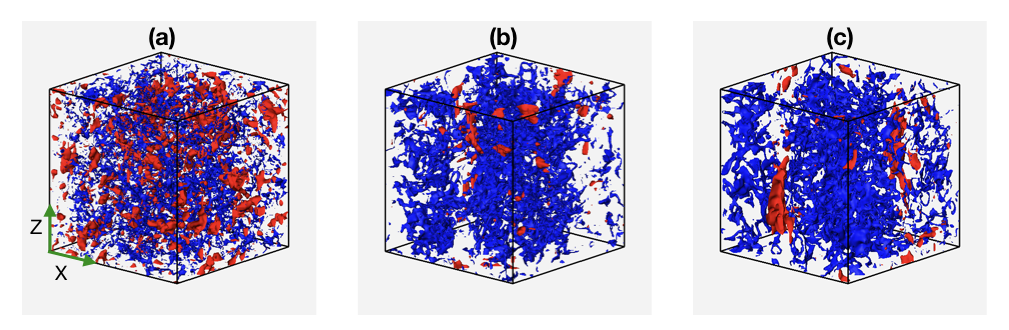}
\caption{ {Volume visualization of isosurfaces of vorticity magnitude ($|\bf \omega| = |\bf \nabla \times \bf v|$) for three different values of rotation rates (a) $Ro = \infty$, (b) $Ro = 0.16$, and (c) $Ro = 0.08$. The blue and red colored isosurfaces represent weak ($|{\bf \omega}| = 0.5$) and intense vorticity ($|{\bf \omega}| = 7$) magnitudes, respectively.}}\label{vor_isosurface}
\end{center}
\end{figure*}

Throughout the duration of all the simulations, presented in this manuscript, the $Re_{\lambda}$ has been kept constant to be approximately 95. The dimensionless Rossby number $Ro \equiv \frac{u_{\rm rms}}{(2 L \Omega)}$ gives a measure of the inertial forces to that of the rotational forces. We chose five different strengths of rotation rates $\Omega = 0$, $0.5$, $0.75$, $1.0$, and $1.5$, yielding corresponding Rossby numbers $Ro = \infty$, $0.24$, $0.16$, $0.12$, and $0.08$. { {The strength of rotation rates $\Omega$ are varied starting from no rotation case to a slow but substantial rotation rate, but restricting ourselves to the region prior to the two-dimensionalization of the flow. In addition, we also wanted to compare our results with prior experimental investigations and hence we used rotation rates comparable to those of reference~\cite{Castello_PRE_2011}.}}

Due to an inverse energy cascade emerging in the presence of rotation, we introduced 
a small frictional term in the form of an inverse Laplacian with a coefficient $\alpha = 0.005$ to arrest the pile up of energy in the smaller modes. The integral length scale given by $ \mathcal{I} = \frac{\Sigma_k E(k)/k}{E}$, where $E(k)$ is the energy spectrum of the flow, changes with the rotation rate due to the emergent inverse cascade. Consequently, the large eddy time $T_{eddy} = \mathcal{I}/u_{rms}$, also decreases with increasing background rotation. Rotation also significantly affects the topology of the flow. The flow regions can be broadly classified into the vortical and straining regions using the ``$\Delta$ criterion"~\cite{chong_1990}, where $\Delta_E$ represents the discriminant of the Eulerian velocity-gradient tensor $\mathcal{A}_{ij} = \partial u_i/ \partial x_j$.  The Eulerian discriminant $\Delta_E$ can also be defined in terms of the invariants, $Q_E$ and $R_E$, of the characteristic equation of $\mathcal{A}_{ij}$ as follows:

\begin{eqnarray}
 {\Delta_E} &=& \frac{27}{4}R_E^2 + Q_E^3, \\
\label{eq:eu_del}
 {Q_E} &=&\frac{1}{2}(-\mathcal{S}_{ij}\mathcal{S}_{ij}+\mathcal{R}_{ij}\mathcal{R}_{ij})\\
 {R_E} &=& \frac{-1}{3}(\mathcal{S}_{ij}\mathcal{S}_{jk}\mathcal{S}_{ki}+ 3 \mathcal{R}_{ij}\mathcal{R}_{jk}\mathcal{S}_{ki})
\label{eq:eu_QR}
\end{eqnarray}

where, $\mathcal{S}_{ij} = 0.5(\partial u_i/\partial x_j + \partial u_j/\partial x_i)$ is known as the strain-rate tensor representing the symmetric component of $\mathcal{A}_{ij}$,  and $\mathcal{R}_{ij} = 0.5(\partial u_i/\partial x_j - \partial u_j/\partial x_i)$ is known as the rotation-rate tensor representing the anti-symmetric component of $\mathcal{A}_{ij}$. The third invariant of the characteristic equation for velocity gradient tensor $\mathcal{A}_{ij}$ is $P = \mathcal{S}_{ii} = 0$, owing to the incompressibility condition. A local measure of the Eulerian $\Delta_E$ becomes an important diagnostic in demarcating the vortical ($ \Delta_E \geq 0$) and the straining ($ \Delta_E < 0$) regions. 

The probability distribution functions (PDFs) of $\Delta_E$ for three different values of rotation are plotted in Fig.~\ref{Dpdf}. In the absence of rotation, clearly, the tail of the PDF is lengthier and appears to be asymmetrically stretched towards the $\Delta_E > 0$ side. The peak of the PDF also appears in the $\Delta_E  > 0$ side. {{This indicates that vortical structures are more prevalent in the flow than straining regions. To ensure this, we computed the skewness of the PDFs at different rotation rates (as shown in the inset of Fig.~\ref{Dpdf}). The skewness of the PDFs always remain positive, though the value decreases with increasing rotation in the system. Excess presence of vortical structures in absence of rotation has also been observed by Perlekar \textit{et al.}\cite{perlekar_2011}}}.
As rotation kicks in, the PDFs of $\Delta_E$ exhibit shorter and more symmetrically stretched tails. Additionally, the skewness of the PDFs of $\Delta_E$ decreases with increasing rotation. Thus, we can conclude that slow rotation in the system suppresses formation of intense vortical or straining regions, while increasing prevalence of weaker regions. {{To clarify this statement, we plot the isosurfaces of weak and intense vorticity magnitude $|\mathbf{\omega}|$ for three different values of rotation rates, as shown in Fig.~\ref{vor_isosurface}. The weak and intense vortical regions are identified by first calculating the mean $|{\mathbf{\omega}}_{mean}|$ and standard deviation $\sigma_{|\bf \omega|}$ from the probability distribution function of vorticity magnitudes obtained at every grid point. Regions of the fluid volume where the vorticity magnitude  $|{\mathbf{\omega}}| < |{\mathbf{\omega}}|_{mean} - 2\sigma_{|\mathbf{\omega}|} $ are classified as weak vortical region, whereas the points having $|\mathbf{\omega}| > |{\mathbf{\omega}}|_{mean} - 2\sigma_{|\mathbf{\omega|}} $ are considered to be in intense vortical regions. The isosurfaces of weak and intense vortical structures plotted in Fig.~\ref{vor_isosurface} correspond to a vorticity magnitude of 0.5 and 7, respectively. In the absence of rotation the structures are more filamented, while with increase of rotation in the system the structures become more elongated, which is a well known effect of rotation. In addition to that, Fig.~\ref{vor_isosurface} also depicts that, slow rotation inhibits the formation of intense vortical region. A similar analysis of the Eulerian discriminant $\Delta_E$ (not shown in here) also points that rotation leads to a suppression of intense vortical or straining regions. Rotation also appears to make the distribution of the vortical and straining regions more homogeneous, indicated by the decrease in skewness of $\Delta_E$ with increasing rotation (as shown in the inset of Fig.~\ref{Dpdf}). Nevertheless, we always observed the prevalence of vortical structures over straining regions, which is reflected in the positive values of skewness of $\Delta_E$.}}

\begin{figure*}[ht]
\begin{center}
\includegraphics[width=0.3\textwidth]{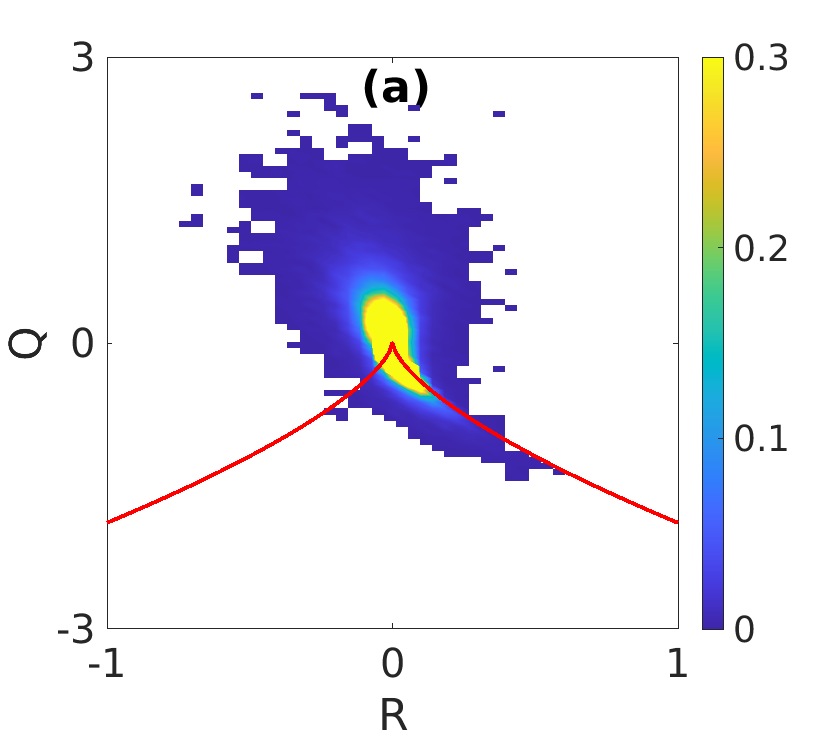} \quad
\includegraphics[width=0.3\textwidth]{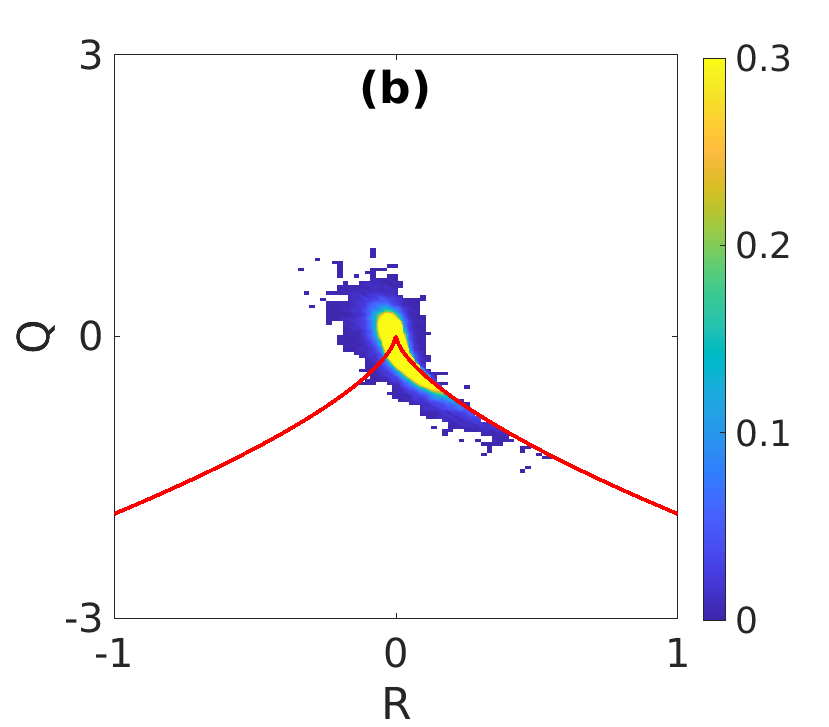} \quad
\includegraphics[width=0.3\textwidth]{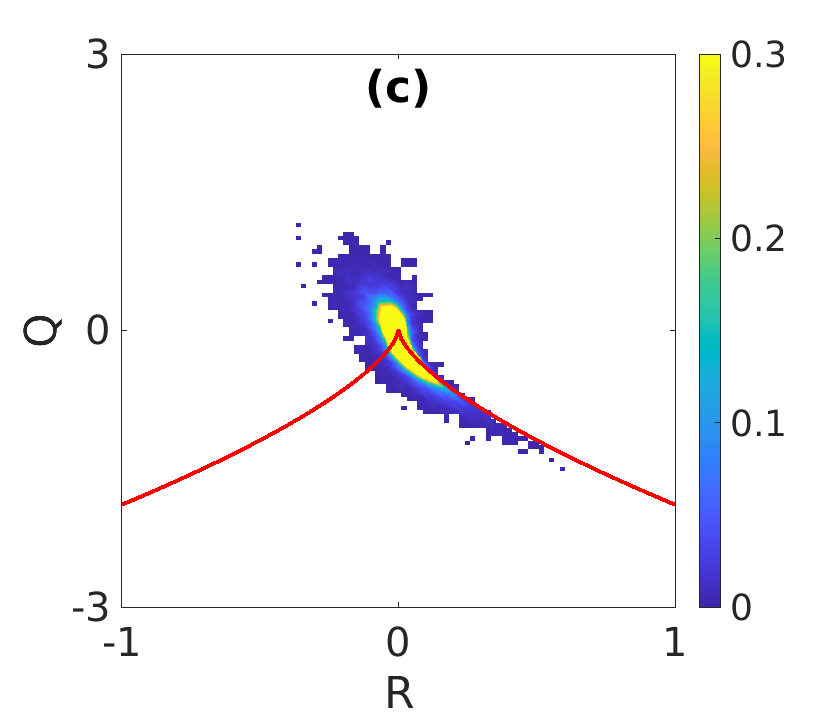} \\
\includegraphics[width=0.3\textwidth]{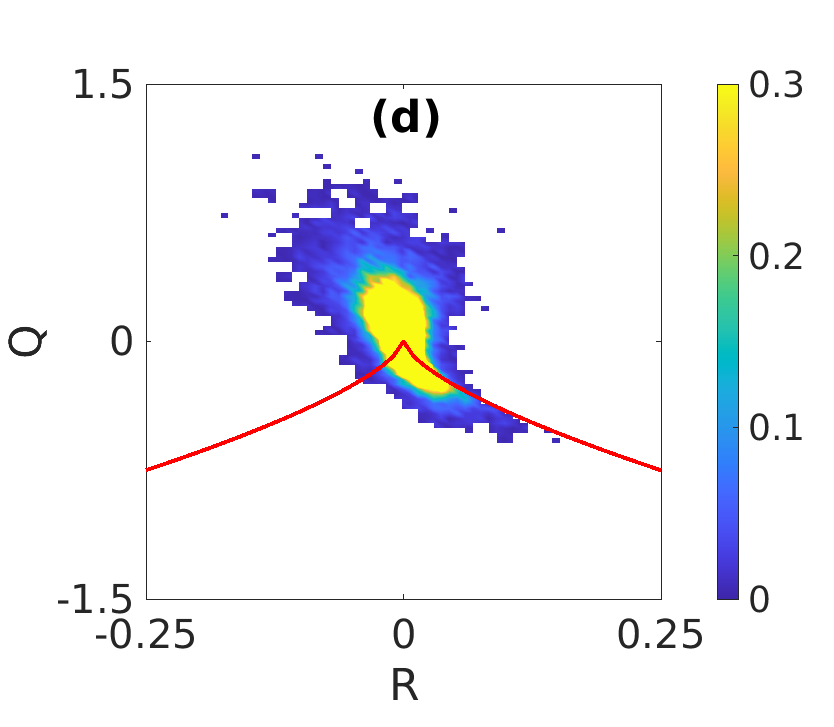} \quad
\includegraphics[width=0.3\textwidth]{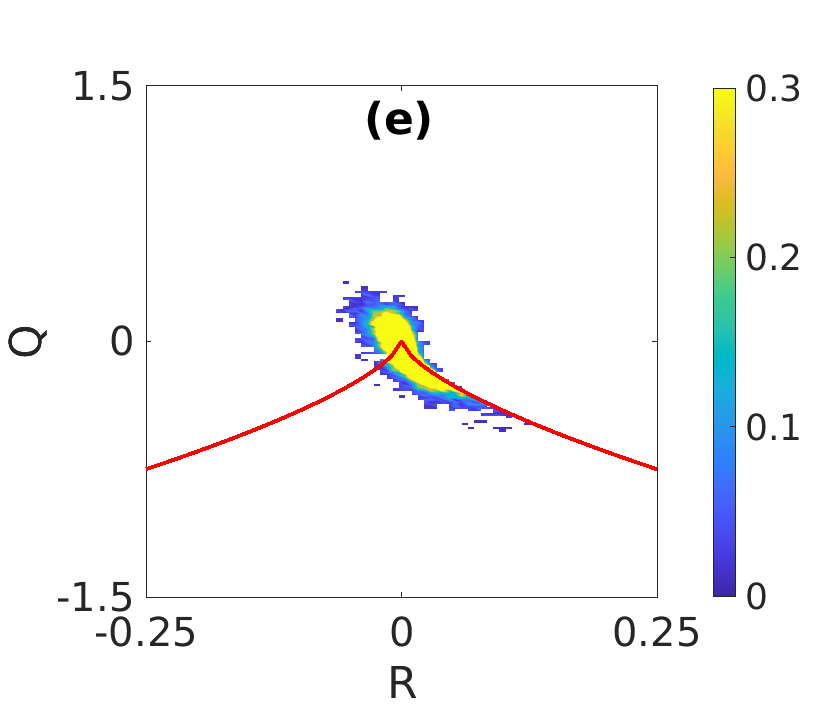} \quad
\includegraphics[width=0.3\textwidth]{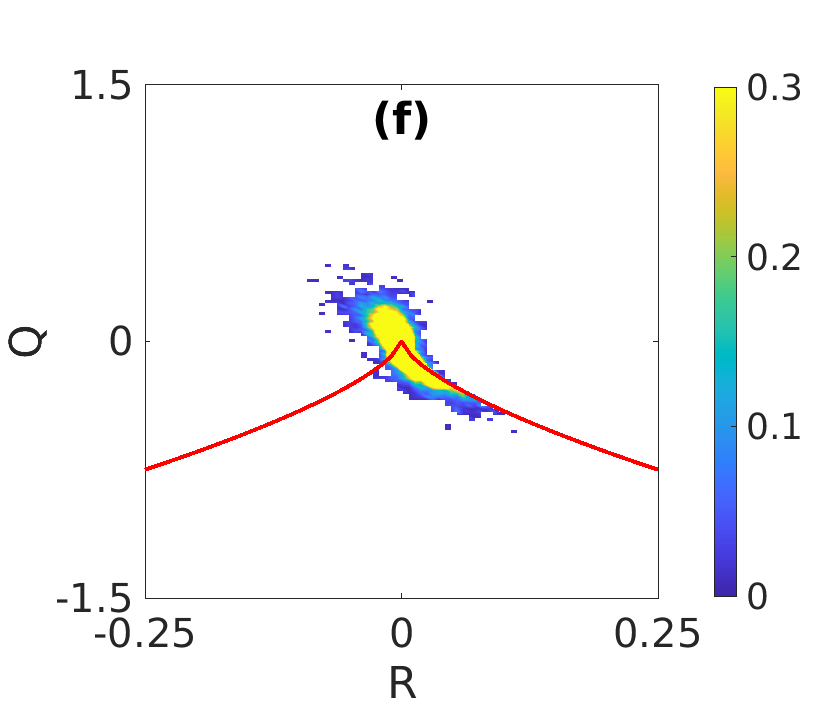} \\
\includegraphics[width=0.3\textwidth]{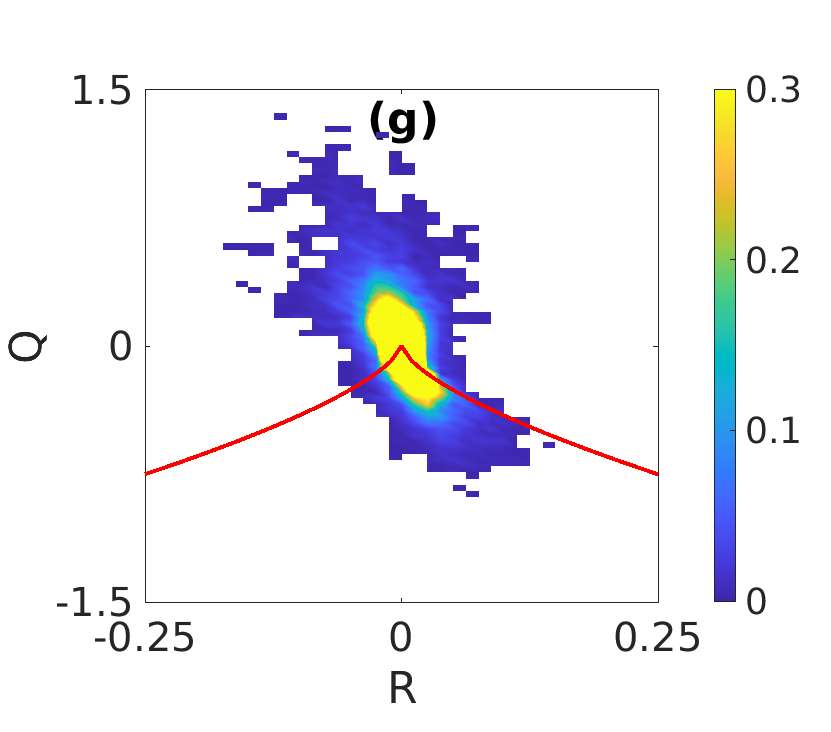} \quad
\includegraphics[width=0.3\textwidth]{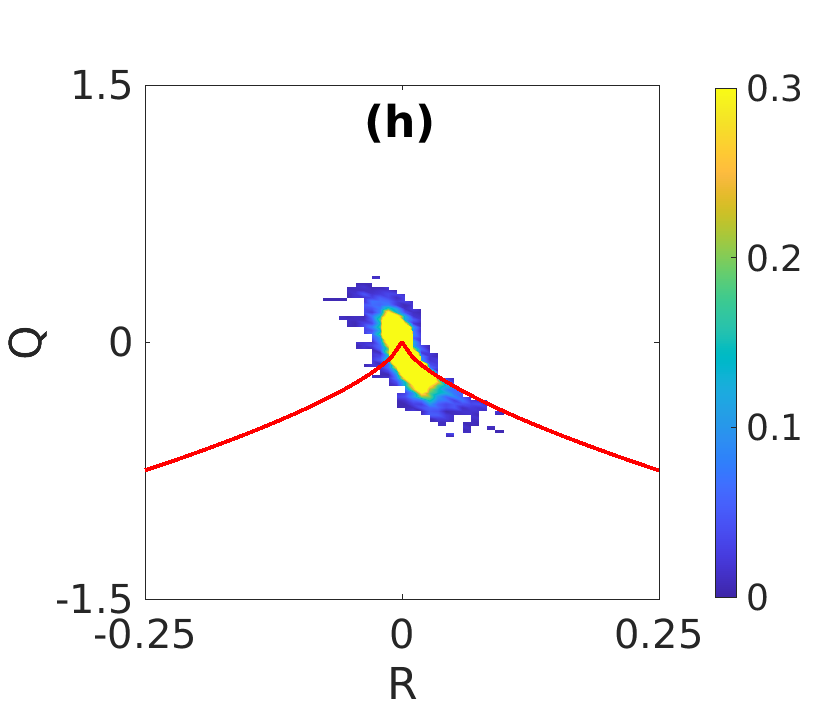} \quad
\includegraphics[width=0.3\textwidth]{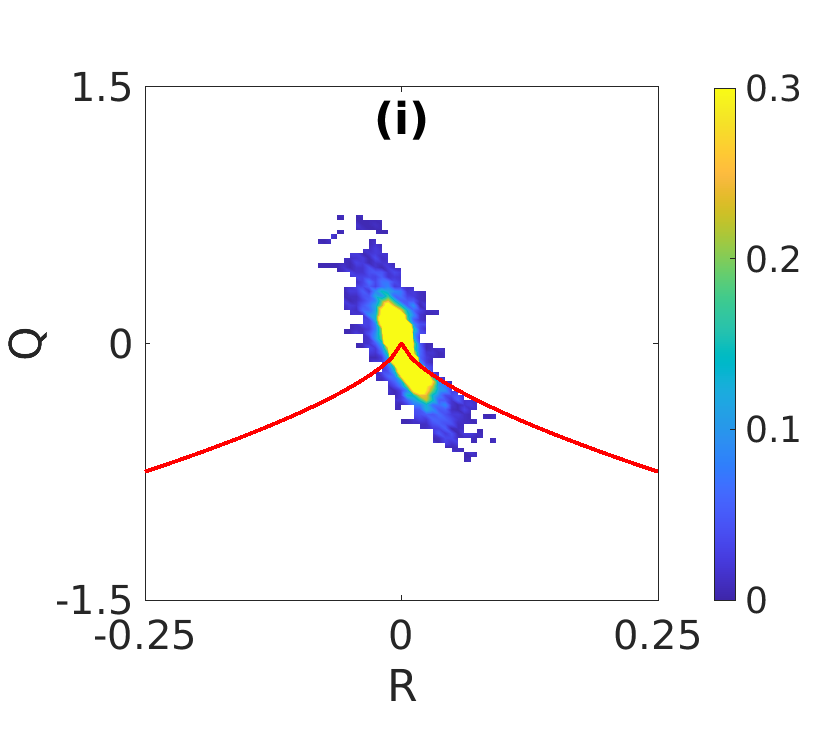} \\
\caption{{{Heat maps}} of joint PDFs of the Lagrangian $Q$ and $R$, calculated along the particle 
trajectories for different values of rotation rates and Stokes number. The parameters are: (a) 
$ Ro = \infty, St = 0$, (b) $ Ro = \infty, St = 0.63$, (c) $ Ro = \infty, St = 1.5$, (d) $ Ro = 0.16, 
 St = 0$, (e) $ Ro = 0.16, St = 0.63$, (f) $ Ro = 0.16, St = 1.5$, (g) $ Ro = 0.08, St = 0$, (h) 
$ Ro = 0.08, St = 0.63$, and (i) $ Ro = 0.08, St = 1.5$. The values of $Q$ and $R$ are non-dimensionalized by the quantity $\chi^2$ and $\chi^3$
respectively, where $\chi = u_{\eta}/\eta$. The solid red curve represents the Eulerian $\Delta_E = 0$ state, which demarcates the vortical 
($\Delta_E \geq 0$) from the straining ($\Delta_E < 0$) region.} \label{JPDF_QR}
\end{center}
\end{figure*}

\begin{figure}
\begin{center}
\includegraphics[width=0.4\textwidth]{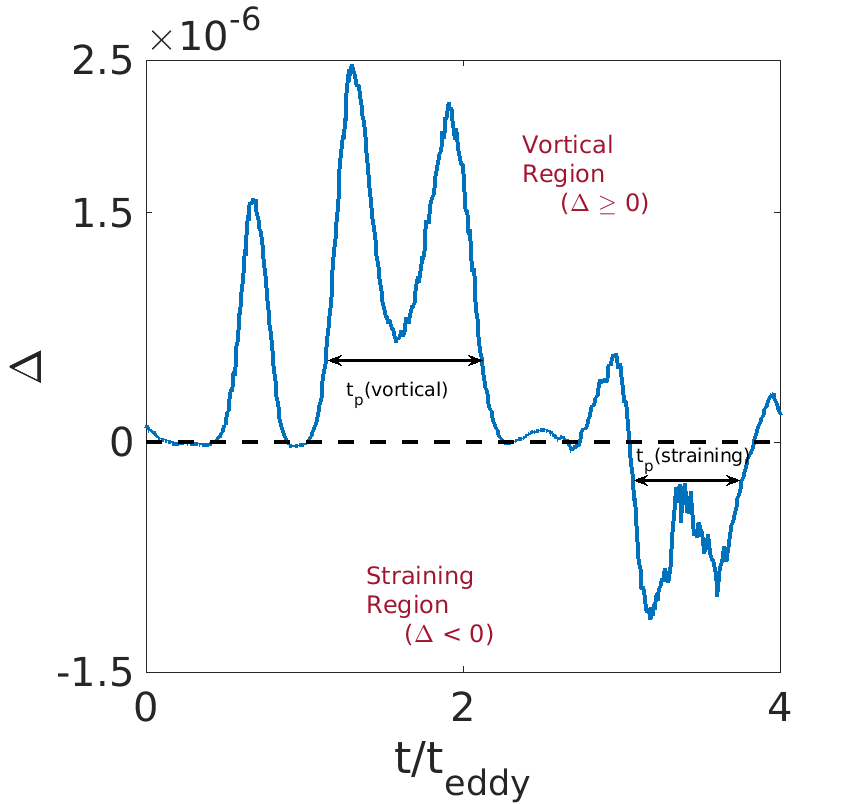}
\caption{{Time series of Lagrangian $\Delta$ (computed as $\Delta_E$ at the particle position) of a random tracer in a fluid with a rotation rate of $Ro = 0.08$. The values of $\Delta$ has been non-dimensionalized by the quantity $\chi^6$ and the time by the eddy turnover time $t_{eddy}$. $\Delta = 0$ is shown by the dashed line, which demarcates the quadrants into vortical ($\Delta \geq 0$) and strain ($\Delta < 0$) dominated regions. The persistence times $t_p$, denote the time duration spent by a particle either in vortical or straining region before getting ejected.}}\label{D_time_series}
\end{center}
\end{figure}
\subsection{Lagrangian particle fields}
Once the flow field attained a statistical steady state, we seeded the flow with $N_p = 10^6$ number of homogeneously distributed mono-disperse particles and allow them to evolve in time. We assumed the particles to be rigid, spherical, and non-interacting, having a radius of $a  \ll \eta$ and density $\rho_p$ much larger than the density of ambient fluid $\rho_f$ ($\rho_p \gg \rho_f$). The equation of motion for the position $\rm \bf x$ and velocity $\rm \bf v$ of the particles can be approximated by modified Maxey-Riley equations~\cite{maxey_riley_1983} in a rotating frame of reference and is given by :

\begin{eqnarray}
\frac{d {\bf x}}{dt} &=&\bf{v}\label{eqn:pv}\\
	\frac{d {\bf v}}{dt} &=& - \frac{({\bf v} - \bf u_{\rm p})}{\tau_{\rm p}} - 2 ({\rm \bf \Omega} 
	\times \bf v) - {\rm \bf \Omega} \times {\rm \bf \Omega} \times {\bf r_p}
\label{eqn:particle_momentum}
\end{eqnarray}
\noindent where  $\bf r_p$ is the position vector of the particle from the axis of rotation and $\tau_p = 2 a^2 \rho_p / 9 \nu \rho_f$ is the relaxation time of the particles with respect to the surrounding fluid flow. The fluid velocity at the particle position $\bf u_{\rm p}$ is computed using a trilinear interpolation scheme to the typically off-grid particle position. We chose 6 different sets of particles with values of inertia quantified by the dimensionless Stokes number $St = \tau_p / \tau_{\eta}$; this allowed us to obtain results for $St = 0, 0.25, 0.63, 0.81, 1.0,$ and $1.5$. We allowed the particles to evolve in time with an integration scheme and time stepping similar to the case of fluid fields. The centrifugal force ejects the particles swiftly out of the box; however, the particles still experience an equivalent flow field due to imposition of periodic boundary conditions. {{The values of $St$ number, beginning with tracers, are chosen to encompass the range where preferential clustering can be observed according to previous studies~\cite{eaton_1994,Bec_2005,Bec_2007_pre,Bec_2007_PRL,Bec_2014}. Proceeding with these parameters, we dedicated ourselves to understand how rotation affects the spatial distribution of particles in the flow and changes the time spent by the particles in different flow regions.}}

\begin{figure}
\begin{center}
\includegraphics[width=0.45\textwidth]{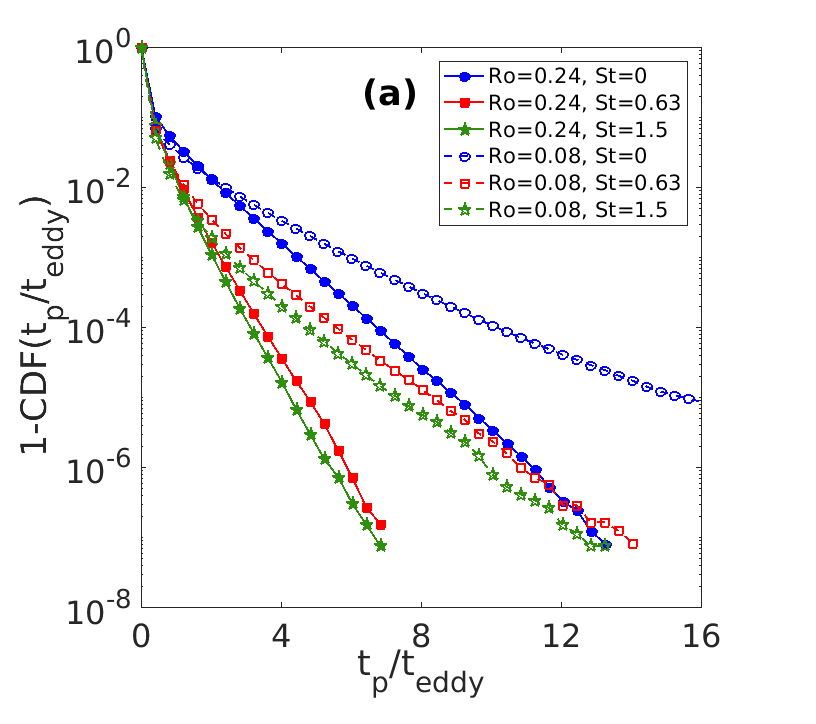}\\
\includegraphics[width=0.45\textwidth]{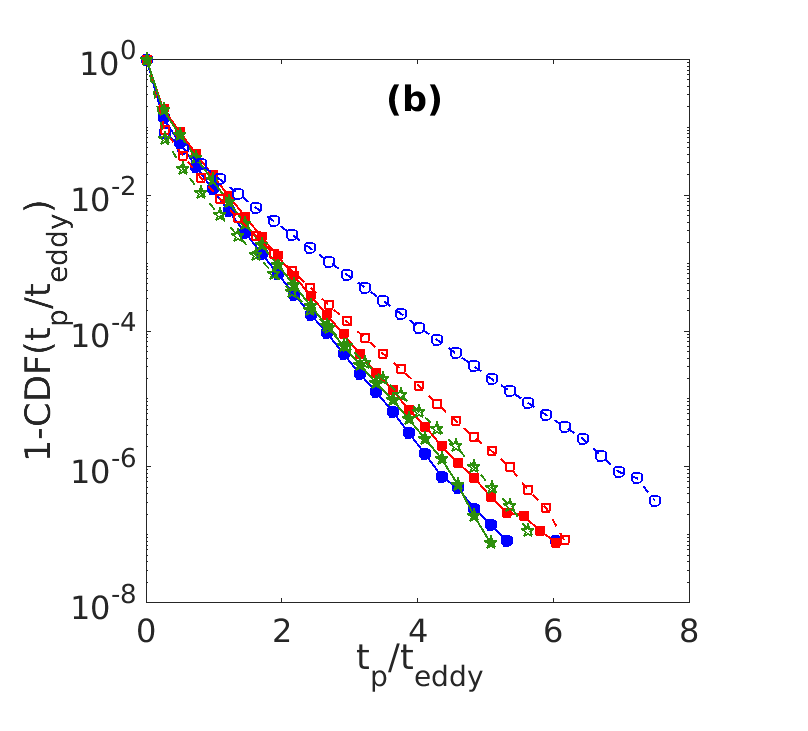}\\
\caption{Cumulative distribution functions (CDFs) of persistence time $t_p$ of particles in (a) vortical
($\Delta \geq 0$) and (b) straining ($\Delta < 0$) regions. The CDFs are plotted for two different 
strengths of rotation ($Ro = 0.24$ and $0.08$) and three different values of Stokes number 
($St = 0, 0.63$, and $1.5$) as shown by the legends.}\label{CDF}
\end{center}
\end{figure}

\section{Results}
\subsection{Distribution of particles in the flow}

We tracked the position and momentum of each particle as they evolved in time. One of the important questions that arise is how do the pseudoforces affect the spatial distribution of particles in the broadly classified vortical or straining regions. For probing this, we computed the instantaneous values of Lagrangian $\Delta$, and the corresponding invariants of the Lagrangian velocity-gradient tensor $Q$ and $R$.  {The Lagrangian values of $\Delta$, $Q$, and $R$  are computed along the Lagrangian trajectories by interpolating the corresponding Eulerian values of $\Delta_E$, $Q_E$, and $R_E$ using a trilinear interpolation scheme to the particle positions ${\bf r}_p(x_p,y_p,z_p)$}. One of the widely used tools to study the trajectories and distribution of the particles in the flow is to plot the {heat maps} of joint probability distribution function (JPDFs) of the Lagrangian quantities $Q$ and $R$. Fig.~\ref{JPDF_QR} shows the {heat maps} of the JPDFs for Lagrangian $Q$ and $R$ for three different strengths of rotation $Ro = \infty$, $0.16$, and $0.08$ and three different values of inertia of particles $St = 0$, $0.63$, and $1.5$. The JPDFs are superimposed on the Eulerian $\Delta_E = 0 $ line (red curve in Fig.~\ref{JPDF_QR}) to demarcate the position of the particles in the flow regions. The area above the red curve ($\Delta_E \geq 0$) depicts the vortical region, and the area below the red curve ($\Delta_E < 0$) depicts the straining region.

In absence of rotation, the JPDFs show the peculiar \textit{teardrop} shape as established by previous studies~\cite{cantwell_1993, Chevillard_2006, Bhatnagar_2016}. The tracers predominantly sample the vortical region, with thin and elongated tails of the JPDFs in the strain dominated region. This is a direct manifestation of the fact that the flow largely consists of vorticity dominated regions; while strain dominated regions are fewer (see Fig.~\ref{Dpdf}). The area of vortical region ($\Delta_E \geq 0$) sampled by heavier particles shrinks, as shown in the JPDFs, due to the phenomenon of preferential sampling. The case of slow rotation $Ro = 0.16$ is equivalent to the non-rotating case; whereas at the highest rotation rate $Ro = 0.08$ the JPDFs deviate from the \textit{teardrop} shape and has more resemblance to that of a kidney bean. This is also directly linked to the change in topology of the flow due to slow rotation, whereby the tracers now sample weaker but still more prevalent vortical structures. In effect, we comprehend that the effect of finite mass ($St$) increases or decreases the area of the \textit{teardrop} structure without altering the shape (depicted in detail by Bhatnagar \textit{et. al}~\cite{Bhatnagar_2016}). On the other hand, rotation in the system, quantified by the Rossby number $Ro$, completely modifies the shape of the $Q-R$ plots in addition to shrinking the area of the plots.

\begin{figure}
\begin{center}
\includegraphics[width=0.45\textwidth]{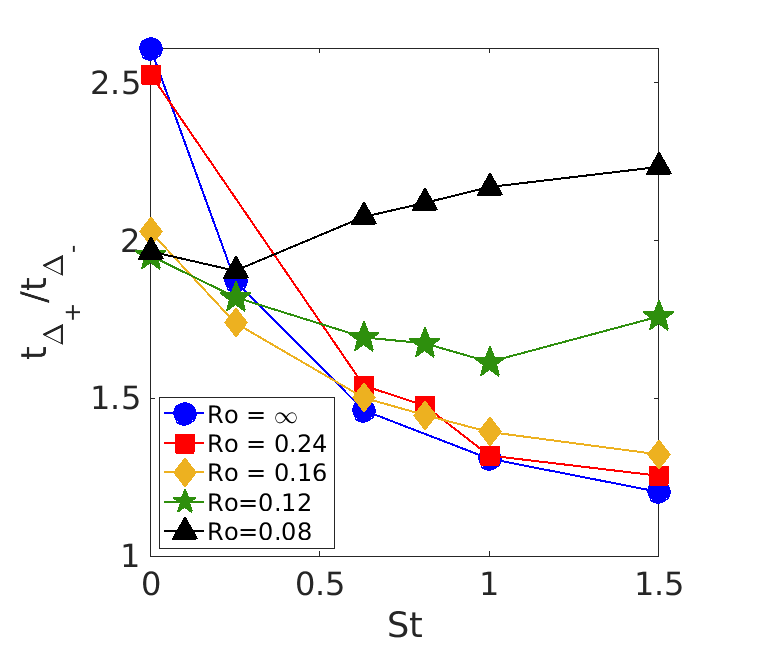}\\
\caption{Ratio of resident times of particles in vortical ($t_{\Delta_+}$) to that in  straining ($t_{\Delta_-}$)
as a function of Stokes numbers $St$. The curves are plotted for five different strengths of rotation $Ro = \infty$, $0.24$, $0.16$, 
$0.12$, and $0.08$ as marked by the legends.}\label{resident_time}
\end{center}
\end{figure}

\subsection{Evolution of Lagrangian trajectories}
We then shifted our focus to temporal statistics of Lagrangian trajectories. Our aim is to investigate how rotation changes the distribution of persistence times  ($t_p$) of particles, the time during which a particle remains trapped either in  vortical or straining region before getting ejected,  in the flow structures. {{To estimate the persistence time $t_p$, we began by analyzing the time evolution of the quantity $\Delta$ along the particle trajectories. The time evolution of $\Delta$, along the trajectory of one such tracer particle for a rotation rate of $Ro = 0.08$, is shown in Fig~\ref{D_time_series}. The persistence times $t_p$ are determined by computing the time durations for which the time signal of $\Delta$ oscillates either in the $\Delta \geq 0$ quadrant (vortical region) or in the  $\Delta < 0$ quadrant (straining region), before changing to opposite sign. Employing this approach, we computed the persistence times of individual particles separately for the vortical and straining regions  (for details, please see references~\cite{perlekar_2011, Bhatnagar_2016}). An equivalent comparison of the persistence times could be the `first-passage times' often used extensively in the statistical Physics literature.}}

{{We, now, compute the PDF of the persistence times $t_p$ separately for particles in the vortical and straining region. The CDFs of the persistence times can be obtained from the PDFs using the relation:
\begin{equation}
    \rm{CDF} (t_p) =  \int_{-\infty}^{t} {P(t_p) dt_p} \\
    \label{eq:CDF}
\end{equation}
\noindent where $P(t_p)$ is the probability that the particle remain in vortical or straining region for a time $t_p$ to $t_p + dt_p$.}}

For maintaining uniformity and comprehensible comparison, we have scaled the persistence times $t_p$ by the large eddy turnover time $t_{eddy}$.{ {We now plot the quantity  ${\widetilde {\rm CDF }} = 1- {\rm CDF}(t_p/t_{eddy})$, which we henceforth term as the shifted CDF, separately for particles in vortical and straining regions for different values of rotation rate and inertia of particles in Fig.~\ref{CDF}.  In three-dimensional turbulence, the shifted  CDFs are known to possess exponentially decaying tails following $\widetilde {\rm CDF}(t_p/t_{eddy}) \propto {\exp}(-t_p/t_{\Delta})$ for both the vortical and straining regions~\citep{Bhatnagar_2016}.}} The fitting parameter, $t_{\Delta}$, gives an estimate of the average resident time of particles in flow regions. In case of two-dimensional turbulence, the tails of the shifted CDFs of Lagrangian persistence time show a power law scaling in the vortical region~\cite{perlekar_2011}, {{while the exponential decay is still preserved for particles spanning the straining regions}}. In our case, at the highest value of rotation rate $Ro = 0.08$, the tails of the CDF in the vortical region show minute deviation from exponential decay, {{owing to the fact that the effects due to rotation, such as stretching of vortical structures and two dimensionalization of flow structures, start emerging at this value of rotation. Nevertheless, the rotation in the system is not strong enough to observe a power law scaling as in reference~\cite{perlekar_2011}}}.

We then fitted the tails of the CDFs with an exponential function to obtain a rough estimate of resident times of particles in vortical ($t_{\Delta_+}$) and in the straining region ($t_{\Delta_-}$). The values of resident times for non-rotating case show good qualitative agreement with those tabulated in reference~\citep{Bhatnagar_2016}. For drawing a more comprehensible conclusion, we plotted the ratio of the quantities $t_{\Delta_+}$ and $t_{\Delta_-}$ as a function of Stokes number $St$ (see Fig.~\ref{resident_time}). Tracer particles, of course, have larger residence times in the vortical region,{{ a direct consequence of the excess presence of vorticity dominated region in the flow (also seen in the $Q-R$ plots of Fig.~\ref{JPDF_QR})}}. At the smallest values of rotation, the tracers spend almost 2.5 times more time in the vortical region than in the straining regions. {{The ratio of the resident times $t_{\Delta_+}$/$t_{\Delta_-}$, for the tracers, dive to 2.0 as rotation rate is increased due to the gradual homogenization in distribution of vortical and straining regions (also seen in the decreasing skewness of PDFs of ${\Delta_E}$, in the inset of Fig.~\ref{Dpdf}).}}

Owing to preferential concentration, this ratio $t_{\Delta_+}$ and $t_{\Delta_-}$ decreases drastically in case of heavier particles {{for moderate values of rotation rates $ 0 < {Ro} \leq 0.12$. The massive particles, now, tend to eject out from the vortical regions.  However, the dominance of vortical structures in the flow makes it non-viable for the massive particles to entirely evade the vortices.  Intuitively, from Fig.~\ref{resident_time}, the ratio $t_{\Delta_+}$/$t_{\Delta_-}$ appears to be determined by a conflicting dominance of rotation rate $Ro$ and Stokes number $St$. For  $Ro \leq 0.16$, the ratio appears to attend a constant. But due to lack of data beyond ${St} > 0.15$, we can not affirm whether the  ratio will increase further or attain a constant value.  In the case of moderate rotation, $Ro = 0.12$, the ratio $t_{\Delta_+}$/$t_{\Delta_-}$ shows less drastic variation, accompanied by a dip at ${St = 1}$. At the highest values of rotation rates, with less intense but more stretched and extended vortical structures, the ratio shows a mild dip at ${St = 0.25}$ and then increases monotonically with $St$. The behavior of the curve at the highest rotation rate is counterintuitive and might be due to  error generated in the fitting parameter due to the deviation from the exponential decaying tails.}}

\begin{figure}
\begin{center}
\includegraphics[width=0.45\textwidth]{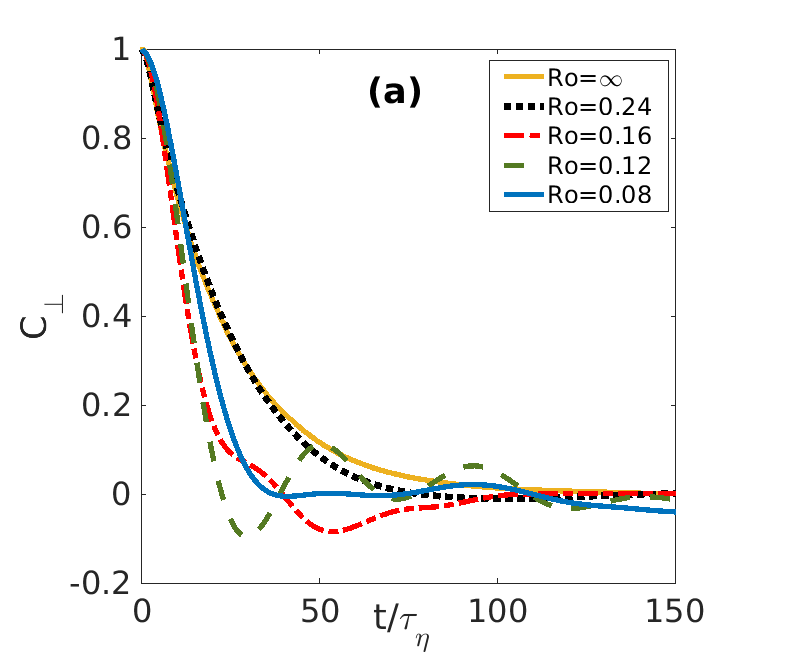}\\
\includegraphics[width=0.45\textwidth]{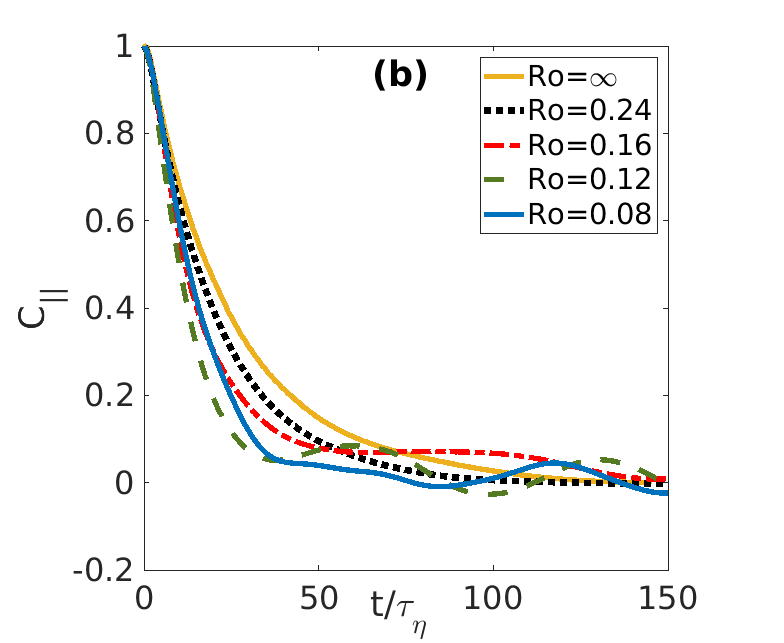}\\
\caption{Autocorrelations of Lagrangian velocities of tracers in the plane (a) perpendicular and (b) parallel
to the axis of rotation as a function of time. The plots are for five different values of rotation rates 
$Ro = \infty, 0.24, 0.26, 0.12$, and $0.08$.}\label{vel_corr}
\end{center}
\end{figure}
 
\begin{figure}
\begin{center}
\includegraphics[width=0.45\textwidth]{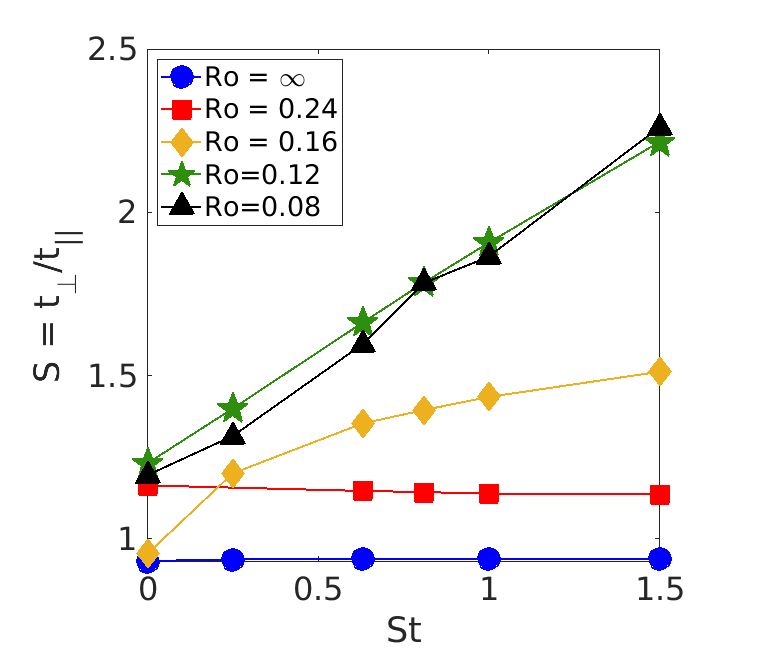}\\
\caption{ Lagrangian anisotropy ratio $S$  as a function of the Stokes number $St$, plotted for five different values to rotation rate (see legends).}\label{fig:S}
\end{center}
\end{figure} 
 
 \subsection{Lagrangian velocity correlations}\label{subsec:vel_corr}
 {Another perspective of investigating Lagrangian trajectories can be presented by computing the corresponding velocity autocorrelation functions.} The $i$-th component of the Lagrangian velocity autocorrelation function can be defined as:

\begin{equation}
{C_i(t) = \frac{\langle v_i(t_0) v_i(t+t_0)\rangle}{\langle v_i^2(t_0)\rangle}}\label{autoc}
\end{equation}

{{Previous investigations~\cite{yeung_pope_1989, Marko_Virant_1997, mordant_2001, Castello_PRE_2011} report that one-component velocity autocorrelation function evolves with a characteristic time of the order of energy injection scale, and the decrement can be modelled by an exponential function. The exponential decay of velocity autocorrelations has been experimentally verified for both non-rotating~\cite{mordant_2001} and rotating turbulence~\cite{Castello_PRE_2011}.
The associated integral time scale can be given as:
\begin{equation}
 T_i = \int\limits_0^{\infty} C_i(\tau) {\mathrm d}\tau\label{aniso_time}
\end{equation}

\noindent and may be interpreted as the time for which a particle remains trapped in a large coherent structure. The integral timescale can be obtained from the autocorrelation function by extracting the fitting parameter of the exponentially decaying tails (as shown in reference~\cite{mordant_2001,Castello_PRE_2011}). A measure of anisotropy, manifested by a rotating turbulent system, can be reckoned by comparing the integral time scales parallel and perpendicular to the direction of rotation (as demonstrated in by Castello and Clercx~\cite{Castello_PRE_2011}). Our primary aim is to probe the effect of background rotation on the velocity autocorrelation functions and associated anisotropy as the Stokes number is varied.}}

{{Fig.~\ref{vel_corr} shows Lagrangian velocity autocorrelation of  tracers for components perpendicular ($C_{\perp}$) and parallel ($C_{||}$) to the axis of rotation. In accordance with previous studies, we fitted the decrement of autocorrelation functions by an exponential decay ${C_i(t)} \propto \exp{(-t/t_0)}$ till the value dropped to $1/e$. The correlations appear to decay faster with increasing rotation rates and can be attributed to the diminishing intense vortical regions, which embodies coherent vortical structures. Our observations, however, are contrary to data of reference~\cite{Castello_PRE_2011}, where an opposite trend has been observed. We have no clear explanation for this discrepancy, except to anticipate that this might be due to the choice of different boundary conditions in the studies. Our simulations assume periodic boundary condition, while the experimental results of reference~\cite{Castello_PRE_2011} enforces rigid wall boundaries. The  exponentially decaying tails become accompanied by oscillations around zero mean, an artifact of the imposed periodic boundary condition accompanied by the effects due to rotation. For heavier particles, the correlations decay much faster and the wiggles appear sooner (not shown in the manuscript), as the centrifugal force ejects the particles more swiftly from coherent vortices. This may lead to the failure of assumption of exponential fitting of the decaying tails.}}

{{We thereafter obtain the integral time scales for the plane parallel ($t_{||}$) and perpendicular ($t_{\perp}$) to the axis of rotation, by separately computing the fitting coefficient ($t_0$) to the functions $C_{\perp}$ and $C_{||}$. The ratio $S = {t_{\perp}/t_{||}}$, now, gives a measure of anisotropy in the system. A plot of the anisotropy measure $S$ as a function of the Stokes number, for various  rotation rate, is given in Fig.~\ref{fig:S}. On account of isotropy in the system in absence of rotation, for all values of $St$, the quantity $S$ ineluctably remains 1. At a low but finite value of rotation ($Ro = 0.24$), the quantity S attains a constant value greater than 1, indicating the onset of anisotropy in the system. The Lagrangian anisotropy seems to be more pronounced for higher $Ro$ and high Stokes numbers $St$. For the mid-range rotation range of $Ro = 0.16$, the anisotropy increases monotonically with the Stokes number and may be projected to attain a constant. At this value of rotation, the particles, owing to their mass, experience the centrifugal force differently. The ratio $S$ increases almost linearly with Stokes number for $Ro \geq 0.12 $; this ambiguous linear increase may have resulted due to the ill-fitted exponential functions to the heavily oscillating velocity autocorrelations for higher values of $Ro$ and $St$. From the perspective of fixed Stokes number, we found that the quantity $S$ always increases monotonically with $Ro$, except for the case of tracers where a slight dip is found at $Ro = 0.12$.}}

\section{Summary and Conclusions}
We investigated the Lagrangian trajectories of heavy particles in rotating turbulent flows. The {heat maps} of JPDs show that the effect of rotation distorts the usual \textit{teardrop} shape of the $Q-R$ plots to a kidney-bean shape. On the other hand, the effect of Stokes number is to alter the area of the $Q-R$ plots while maintaining the \textit{teardrop} shape. We also computed the cumulative distribution functions of the persistence time $t_p$, to get a rough estimate of the average resident time of particles in the vortical or straining region. For lower values of rotation, the tails of the CDFs show a good fit to an exponential shape. However, at the highest rotation rate, corresponding to $Ro$ = 0.08, the tails of the CDFs show a deviation from a perfect exponential shape. Additionally, we also verified that at the highest rotation rates, the tails of the CDFs neither follow a power law nor a stretched exponential. The average resident times of the heavy particles depends strongly on both the rotation changing the topology of the flow, and the inertia of the particle. {{We further calculated the velocity autocorrelation function and the fitting parameter gave a rough estimate of the Lagrangian integral times w. r. t the plane parallel ($t_{||}$) and perpendicular ($t_{\perp}$) to the axis of rotation. The asymmetry ratio $S$ gives an estimate of the Lagrangian anisotropy in the system for various values of $St$ number and rotation rates. The anisotropy is more pronounced in the case of high rotations and high Stokes number. We were not able to furnish any solid explanation for the linear growth of Lagrangian anisotropy $S$ with increasing Stokes number $St$. But we anticipate that this may be due to the erroneous fitting of the velocity correlations due to the early onset of oscillations in the correlation functions.}}

A  detailed investigation for the spatial distribution by computing the correlation dimension $\mathcal{D}_2$ can follow as an immediate extension of this work.  The Lagrangian anisotropy also exhibits non-monotonic behavior with both $St$ and $Ro$, and depends on a complex competition of both the parameters. The dip in the fitting parameter $t_{\perp}$ near mid-range rotation rate remains an open problem. This present study could also be extended for cases of high rotation rates leading to geostrophic range of flows and flows subjected to shear forces. Additionally, the effect of rotation on self-interacting particles is left as a future scope of study.

\section{Acknowledgements}
The author is grateful to Prof S. S. Ray and Prof. R. Govindarajan of ICTS-TIFR, India for their encouragement and expert advice during the entire course of investigation. The simulations were performed at the MOWGLI and TETRIS clusters of ICTS-TIFR India. The author would also like to thank Prof. J. Schumacher, Dr. M. Brynjell-Rahkola, and Mr. V. Pushenko from TU Ilmenau, Germany, and Dr. V. Valori from ETH Zurich, Switzerland for important discussions and comments during preparation of the manuscript. 

\section{Funding}
The author is supported by the grants SCHU 1410/29-1 and SCHU 1410/30-1 of the Deutsche Forschungsgemeinschaft (DFG).

\bibliographystyle{apsrev4-1}.
\bibliography{ref}
\end{document}